\documentclass[preprint,showpacs,preprintnumbers,amsmath,amssymb,epsfig]{revtex4}

\usepackage{epsfig}
\usepackage{subeqnarray}

\newcommand{\beq}{\begin{eqnarray}}
\newcommand{\eeq}{\end{eqnarray}}
\newcommand{\bsubeqa}{\begin{subeqnarray}}
\newcommand{\esubeqa}{\end{subeqnarray}}

\newcommand\ket [1]{\vert #1 \rangle}

\newcommand\mtx [3]{\langle #1 \vert #2 \vert #3 \rangle}

\newcommand\Dw {\Delta \omega}

\begin{document}

\title{Coherent Control of spin-orbit precession with shaped laser pulses}
\author{B\'eatrice Chatel, Damien Bigourd, S\'ebastien Weber, and Bertrand Girard$^*$}
\affiliation{
Laboratoire de Collisions, Agr\'egats, R\'eactivit\'e \\
(CNRS, Universit\'e de Toulouse-UPS), IRSAMC, Toulouse, France}
\date{\today}

\begin{abstract}

Spin precession in Rubidium atoms is investigated through a pump-probe technique. The excited wave packet corresponds to a precession of spin and orbital angular momentum around the total angular momentum. We show that using shaped laser pulses allows us to control these dynamics. With a Fourier transform limited pulse, the wave packet is initially prepared in the bright state (coupled to the initial state) whereas a pulse presenting a $\pi $ step in the spectral phase prepares the wave packet in the dark state (uncoupled to the initial state).

\end{abstract}
\maketitle

Atomic fine structure is well understood. It has been widely studied through various spectroscopic techniques, including quantum beats which allow to measure directly the splitting without the limitation due to the Doppler effect \cite{Haroche_Quantum_beats74}. More recently, several time-resolved techniques have allowed to observe in real-time the temporal evolution of fine structure doublet states in alkali atoms. This temporal evolution corresponds to the precession of the spin and angular momentum around the total angular momentum \cite{Rozmej_precession94}. It was
first indirectly observed in two pulse interferometry experiments (temporal Coherent Control)
\cite{Snoek96Rydbergstates, Blanch97Cs, Bouchene98KCs, Bouchene99, BoucheneOC00} : The two-path interference reveals the absorption lines exactly as in Ramsey fringes or in Fourier transform spectroscopy. The beat between two transitions is related to the exited state structure only if they share the same initial state which cannot be proven unless in the saturation regime \cite{Nicole99a, Nicole_JMO02}. Counter examples can for instance be found in room-temperature molecules \cite{Scherer92Rotation,Blanch95, Blanch98Cs2,Ohmori03}.
More recently, standard pump-probe experiments have been performed: The ionization probability of the excited state depends on the orientation of the orbital angular momentum with respect to the probe laser polarization \cite{Zamith2000a,Sokell00}. This reveals directly the orbital angular momentum precession. Alternatively, this precession can also be observed by direct photoelectron angular distributions \cite{Leone_JPB06} or
Half Cycle Pulses whose interaction with the excited atom depends on the orbital angular momentum orientation \cite{Conover01_Rydberg_QB, Bucksbaum_PRA06spinorbit}. In a less expected way, this angular momentum precession produces strongly contrasted interferences in ladder climbing with chirped pulses when the ladder intermediate step is a fine structure doublet \cite{chatelNa03PRA, chatel04PRA,Merkel07PRA1}. A sequence of two chirped pulses has also been used to excite selectively one of the two excited states \cite{Sauerbrey02,Sauerbrey03}.

The possibility of applying spin-orbit precession to the production of spin polarized electrons \cite{Bouchene01,Nakajima_spin_polarization} or nuclei \cite{Nakajima_PRL07nuclearspins} has been discussed. Similarly, it was recently proposed to apply spin-orbit precession to the measurement of attosecond pulses \cite{Itakura_PRA07}.

Pulse shaping techniques \cite{weiner-slm-1990,Dazzler97,weiner00} have proven to be a strongly versatile tool to alter and control the ultrashort dynamics of quantum systems \cite{Goswami03Review, Dantus04review, Fielding_Review05}. These have been used in optimal control strategy within a closed loop approach \cite{Judson92} in which the pulse shape is optimized in order to reach the predefined goal \cite{Assion98, RabitzScience2000, Levis01, Motzkus02bio, woste-co-sience2003}. On the other hand, an open loop approach can be used to demonstrate the effect of particular pulse shapes on the transition probability or on the dynamics of the system \cite{silberberg98, silberberg99, Motzkus00, silberberg01, zamith01, Degert02CTshaped, silberberg02, Amitay01shift, Leone02WPOptimization, Leone02a, Leone03, Degert03NaI, Dantus03multiphoton, RbShapingAPB04, dudovich_momentum_prl_2004, renard2004, Faucher_PRA05Pulseshaping, KatsukiScience06, KatsukiPRL06, OhmoriPRA07, MonmayrantCT-reconstruction-05, MonmayrantCT-spirograph-06, MonmayrantCT-spirograph-optCom06}.

In this paper we present a pump-probe experiment performed in the fine structure levels of the (5s - 5p) transition in atomic Rubidium performed either with a Fourier transform limited (FTL) pump pulse or with a shaped pump pulse. We first describe the precession dynamics of the spin -orbit wave packet, using the uncoupled basis set. In this framework, the spin is spectator during the excitation so that the wave packet is initially created in a state with the same spin orientation. The natural evolution corresponds to the precession of the spin and orbital angular momenta around the total angular momentum. This precession is induced by the spin-orbit coupling. After half a period, the spin has flipped. The atom is in the dark state, uncoupled to the initial state. We then study the excitation by a shaped laser pulse in which a $\pi $ phase step has been applied between the two absorption lines. We show that the system can then be directly excited to the dark state (with the spin flipped).

\section{Theory}
\subsection{Fine-structure states}
Fine structure effects are due to relativistic interactions and they result in the splitting of atomic levels. The main consequence is the coupling between the spin $\vec{s}$ and orbital $\vec{l}$ electronic angular momenta, so that the energy levels are eigenstates of the total angular momentum $\vec{j}$. Hence, depending on the required energy scale, atomic levels can be either described in the uncoupled basis set  $|n,l,m_l,s,m_s>$ or in the coupled basis set $|n,(l,s);j,m_j>$. In this paper, we will use the simplified notations $|l,m_l,m_s>$ and $|l,j,m_j>$ for these two basis sets respectively. For fine structure states associated to a $\textrm{P}$ state ($l=1$) with $s=1/2$ , we have $j=1/2$ and $j=3/2$. Both basis sets can be used to describe the ultrashort dynamics of these fine structure levels excited by an ultrashort laser pulse \cite{Zamith2000a,Sokell00,Bucksbaum_PRA06spinorbit}. In the coupled basis set (the actual energy levels of the atom), the initial state prepared by the ultrashort pulse appears as a coherent superposition of the states with $j=1/2$ and $j=3/2$. On the other hand, in the uncoupled basis set, and in the limit of laser pulses much shorter than the internal dynamics within the fine structure levels (equal to the reciprocal of the energy splitting), the initial state is equal to the state with the same spin orientation ($m_s$ value) as the ground state.

\subsection{Temporal evolution of a two-level system after short pulse excitation}
In the weak field regime, the temporal evolution of the excited state wave function is given by first order perturbation theory:
\beq \label{psi1}
\ket{\psi_e (t)} = \frac {i}{\hbar} \sum\limits_k {\int\limits_{-\infty}^t \, {{\cal E} (t') \, \mu_{kg} \, e^{i \omega_k (t'-t)} \, dt' \, \ket{k}}}
\eeq
where $\ket{k}$ is an excited state of energy $E_k= \hbar \omega_k$, $\mu_{kg}=\mtx{k} {\mu}{g}$ the transition dipole moment from the ground state $\ket{g}$. The summation over $k$ involves all the possible excited states. The electric field, polarized along $Oz$, can be written in the Rotating Wave Approximation (RWA)
\beq \label{field}
{\cal E} (t)= {\cal E}_0  f(t) \, e^{-i \omega_L t}
\eeq
where $f(t)$ is the dimensionless temporal shape. Its characteristic duration is $\tau_L$, and its spectral width is $\Dw _L$ so that $\pi / \Dw _L$ is its shortest temporal feature.
At the end of the exciting laser pulse ($t \ge \tau_L$) the wave function is
\beq \label{psi2}
\ket{\psi_e (t)} = \Omega \tau_L \sum\limits_k {a_{k} \, e^{-i \omega_k t} \ket{k}}
\eeq
where $\Omega$ is the generalized angular Rabi frequency defined by
\beq
\left| \Omega \tau_L \right|^2 = \frac {1}{\hbar^2} \sum\limits_k {\left|{\cal \tilde{E}} (\omega_k)\mu_{kg}\right|^2 }
\eeq
 so that the $a_k$ coefficients are normalized ($\sum\limits_{k'} {\left| a_{k'} \right|^2} = 1$):
 \beq  \label{ak}
 a_{k}= \frac {{\cal \tilde{E}} (\omega_k)\mu_{kg}} {\sqrt{\left( \sum\limits_{k'} {\left|{\cal \tilde{E}} (\omega_{k'})\mu_{k'g}\right|^2 } \right)}}
 \eeq
The expression (\ref{psi2}) is valid for any pulse shape and as long as $\Omega \tau_L \ll 1$ in order to be in the perturbative regime. Shaping the pulse changes the phases and amplitudes in ${\cal \tilde{E}} (\omega_k)$ defined as the Fourier transform of ${\cal E} (t)$. Therefore, both the short time (laser driven) and long time (field free) dynamics are modified. Moreover, since the pulse duration is also affected, the transition between these two regimes is also shifted towards longer times. However, it should be noted that the field free evolution is governed by the spectral phases and amplitudes at the transition frequencies only whereas the short-term, laser driven, evolution depends on the full shape \cite{zamith01,silberberg02,Degert02CTshaped}.

We now examine how these general statements apply to a set of two excited states in which our experiment has been carried out. We will briefly discuss in the outlook how this approach can be extended to the case of several excited states.

The field free evolution of a coherent superposition of two levels separated by the energy $ E_2 - E_1 = 2 \hbar \Dw$ is an oscillation at the period $T = \pi / \Delta \omega $ between the bright state (or doorway state) $\ket{\psi_B}$ and the dark state $\ket{\psi_D}$ defined respectively by:
\bsubeqa \label{CUCstates}
\ket{\psi_B}= a_1 \ket{1} + a_2 \ket{2}\\
\ket{\psi_D} = a_2 \ket{1} - a_1  \ket{2}
\esubeqa
where, from now on, $a_1$ and $a_2$ are the coefficients defined by Eq. (\ref{ak}) but corresponding specifically to a Fourier limited pulse.
More precisely, for $t \ge \tau_L$
\beq \label{psi3}
\ket{\psi_e^{FT} (t)} & = & \Omega \tau_L  \, e^{-i \overline{\omega} t} \left( a_1 e^{i \Dw t} \ket{1} + a_2 e^{- i \Dw t} \ket{2} \right) \nonumber \\
 & = & \Omega \tau_L  \, e^{-i \overline{\omega} t} \left[ \left( \cos \Dw t +i (a_1^2 - a_2^2) \sin \Dw t \right)  \ket{\psi_B} + 2i a_1 a_2 \sin \Dw t \ket{\psi_D} \right]
\eeq
where $\overline{\omega} = (\omega_1 + \omega_2)/2$. The probabilities of finding the system in the bright or dark states are respectively:
\bsubeqa \label{Proba:FT}
\hfill P_B^{FT}(t) & = & \left( 1 - 4 a_1^2 \, a_2^2 \sin ^2 \Dw t \right) \Omega^2 \tau_L ^2\\
\hfill P_D^{FT}(t) & = & \left( 4 a_1^2 \, a_2^2 \sin ^2 \Dw t \right) \Omega^2 \tau_L ^2
\esubeqa

In the case of ${\rm P}_{1/2}$ - ${\rm P}_{3/2}$ spin-orbit states excited from a ${\rm S}_{1/2}$ state, one has $\mu_{2g} / \mu_{1g}= \sqrt{2} $. Moreover, with a laser wavelength centered in the middle of the two transitions so that ${\cal \tilde{E}}(\omega_1) = {\cal \tilde{E}}(\omega_2)$, we also have $a_1= 1/\sqrt{3}$ and $a_2 = \sqrt{2/3}$. In this case, the bright state population oscillates between 1 and 1/9 whereas the dark state population oscillates between 0 and 8/9 with an opposite phase. Hence, even with strongly unbalanced probabilities to reach states $\ket{1}$ and $\ket{2}$ (a factor of 2), we have almost complete population transfer between the bright state and the dark state. One should note that although the system is fully in the bright state for $t_p=pT$, it is only partially transferred to the dark state which is maximally populated at $t'_p=(p+1/2)T$. This oscillation has been fully observed in potassium atom \cite{Zamith2000a}. Similarly, it was shown that the roles of the dark and bright states could be inverted by rotating the probe pulse polarization by $90^\circ$ \cite{Sokell00}.

\begin{figure}[!ht]
\begin{center}
\epsfig{figure=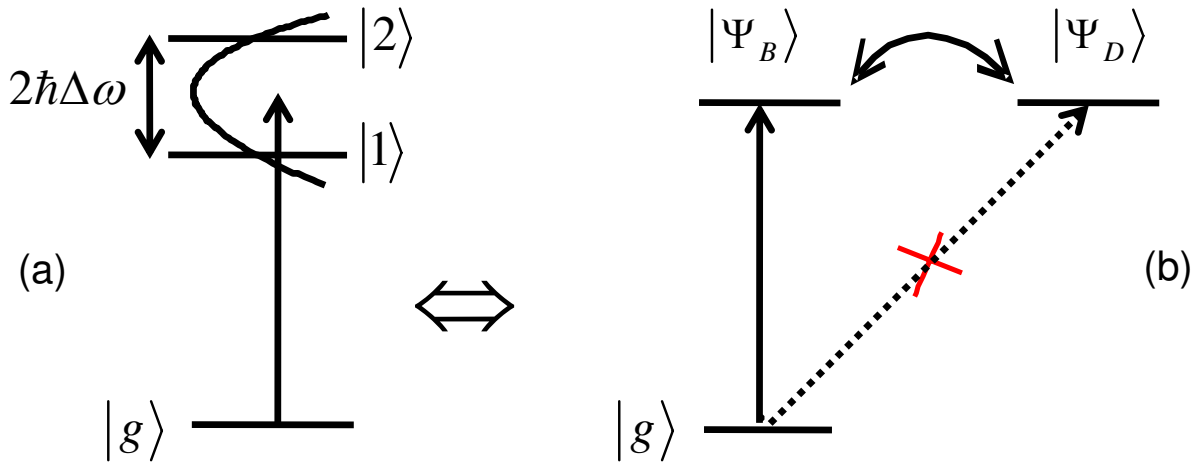,width=0.9\linewidth}
\caption[States]{Excitation scheme (a) in the coupled basis and (b) in
the uncoupled basis.} \label{States}
\end{center}
\end{figure}

Fourier transform limited (FTL) pulses have a flat spectral phase. In the case of a pulse duration much shorter than the field free dynamics ($\tau_L \ll T$), then the system is in the bright state $\ket{\psi_B}$ immediately after the end of the laser pulse and the oscillation starts from this state (see Fig. \ref{States}b). In the case of a longer pulse, then both bright and dark states are populated during the interaction: The bright state is populated first and then there is a continuous flow of population from the bright state towards the dark state while in parallel excitation keeps filling the bright state from the ground state. At the end of the laser pulse, the field free oscillation given by Eq. (\ref{psi3}) takes place and its phase is the same as with a short laser pulse. Indeed, the long term evolution depends only on the spectral phase and amplitude at the transition frequencies and not on the whole spectrum.

With shaped pulses, the field free evolution consists also of an oscillation between two states as long as both stationary states are populated (${\cal \tilde{E}} (\omega_1) \ne 0$ and ${\cal \tilde{E}} (\omega_2) \ne 0$). These two states are in general different from the dark and bright states defined in Eq. (\ref{CUCstates}). We consider here the particular case in which a phase step of $\pi$ is applied at a frequency intermediate between the two transitions and the pulse amplitudes are unaffected. The coefficients of states $\ket{1}$ and $\ket{2}$ are thus $a_1$ and $-a_2$ respectively. In this case, the oscillation takes place between the same two states as with FT limited pulses, but it is however phase shifted by $\pi$. The field free evolution is given by
\beq \label{psi4}
\ket{\psi_e^{sh} (t)} & = & \Omega \tau_L  \, e^{-i \overline{\omega} t} \left( a_1 e^{i \Dw t} \ket{1} - a_2 e^{- i \Dw t} \ket{2} \right) \nonumber \\
 & = & \Omega \tau_L  \, e^{-i \overline{\omega} t} \left[ \left( i \sin \Dw t + (a_1^2 - a_2^2) \cos \Dw t \right)  \ket{\psi_B} + 2 a_1 a_2 \cos \Dw t \ket{\psi_D} \right]
\eeq
which is the same dynamics as the one produced by a FTL pulse but shifted by $T/2$. The population evolutions are given by
\bsubeqa \label{Proba:sh}
\hfill P_B^{sh} (t) & = & \left( 1 - 4 a_1^2 \, a_2^2 \cos ^2 \Dw t \right) \Omega^2 \tau_L ^2 \\
\hfill P_D^{sh} (t) & = & \left( 4 a_1^2 \, a_2^2 \cos ^2 \Dw t \right) \Omega^2 \tau_L ^2
\esubeqa
which corresponds to the same oscillation between the dark and bright states as in the FTL case, but it is here out of phase by $\pi$.

\begin{figure}[!ht]
\begin{center}
\epsfig{figure=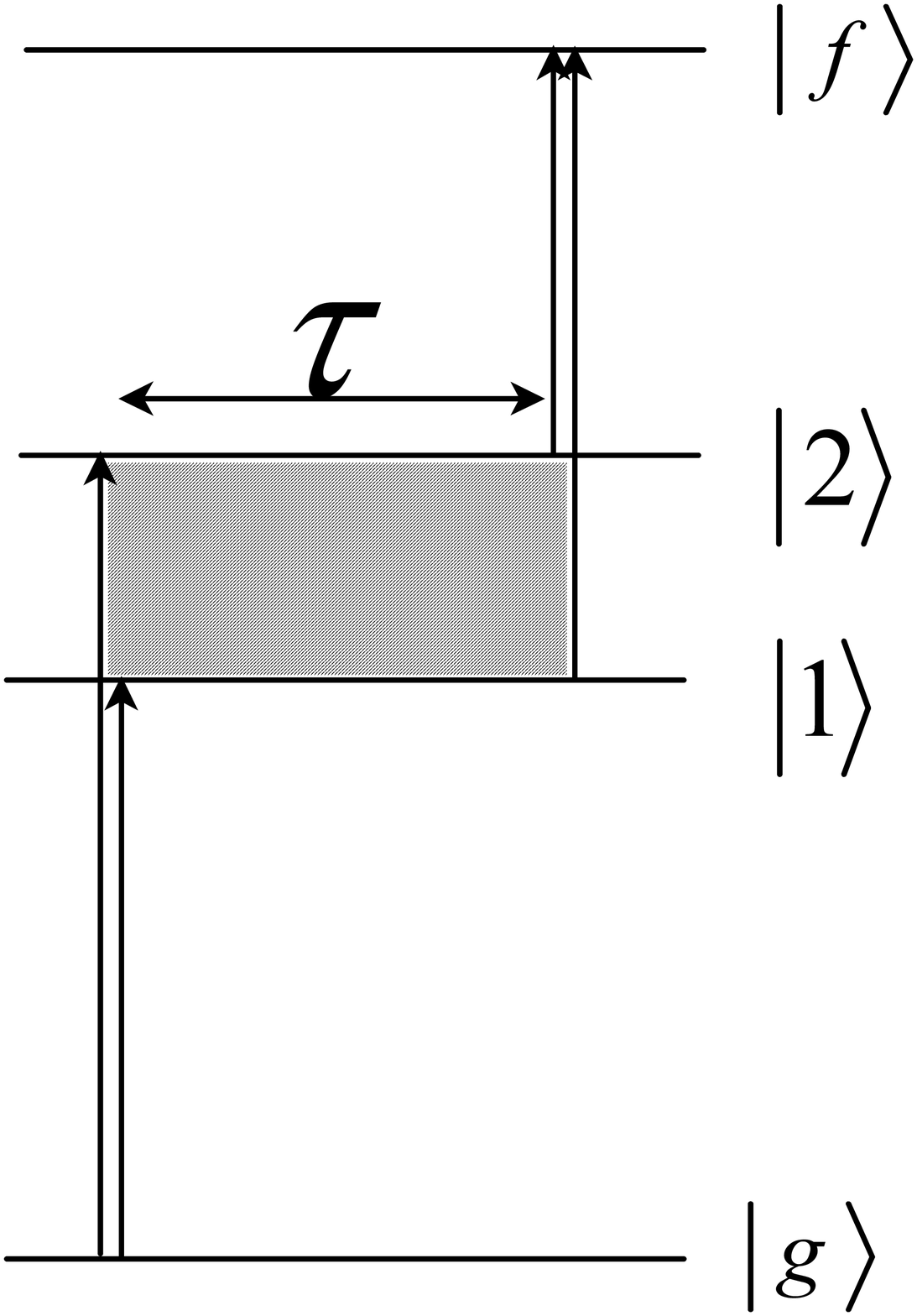,width=0.3\linewidth}
\caption[Ionization]{Detection scheme : Both states can be transferred
towards the same final state using a short probe pulse. Interference produces
oscillation as a function of the pump-probe delay $\tau$.} \label{Ionization}
\end{center}
\end{figure}

In order to observe these oscillations, we use a FTL pulse as a probe which excites the system towards the (8s, 6d) Rydberg states. The requirement to observe oscillations is that the excitation probability should be different for the bright and dark states. This is equivalent to having non vanishing probability excitations from both stationary states $\ket{1}$ and $\ket{2}$ as shown schematically on Figure \ref{Ionization}. The oscillation contrast of the pump-probe signal is in general smaller than the one of the population oscillation, unless one of the two (bright or dark) states has a negligible detection probability as compared to the other.

\begin{figure}[!ht]
\begin{center}
\epsfig{figure=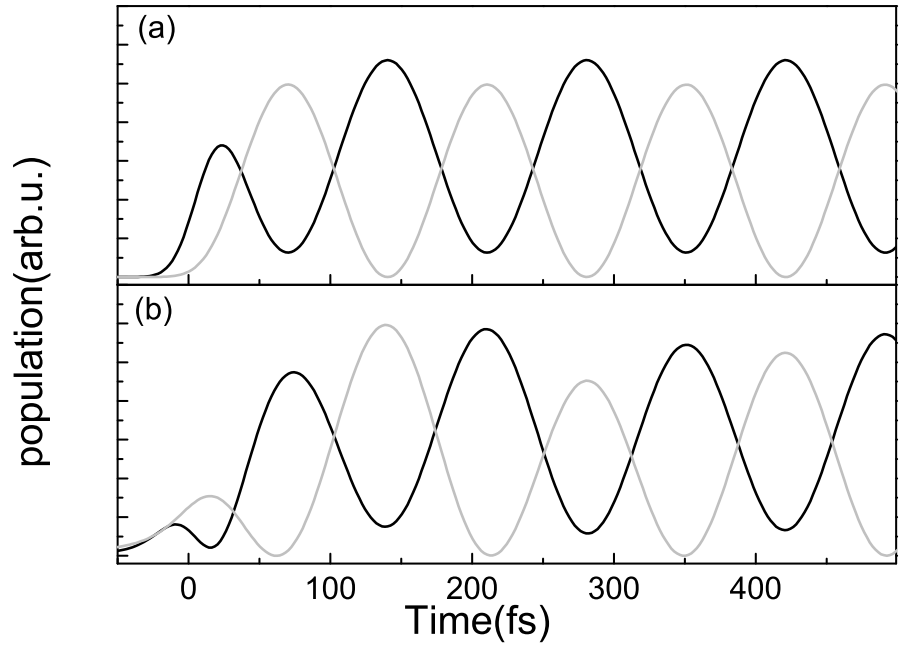,width=0.8\linewidth}
\caption[Theory]{Theory. Temporal evolution of the population in the
bright (black line) and dark (gray line) states for FT limited
pulses (a), for a shaped laser pulse (b).} \label{Theory}
\end{center}
\end{figure}

Figure \ref{Theory} presents the predicted population evolution in the bright state and in the dark state for the FTL case and the $\pi$ jump case for a fine structure doublet. The calculations have been performed starting directly from Eq. (\ref{psi1}). One clearly sees the features discussed above: Strong contrast (89\% modulation depth) of the oscillations, shift of $\pi$ of the oscillations between the bright and dark state populations, and finally from the FTL case to the shaped pulse case. We discuss now the initial evolution during the interaction with the laser pulse. In the FTL case, the bright state is first populated and the dark state becomes populated after half an oscillation period. In the shaped case, the dark state receives a significant population before the bright state is fully populated (after half a period).

\section{Experimental set-up}
\label{setup}

The experimental set-up is a standard pump-probe experiment. 
The 5s - 5p ($^2$P$_{1/2}$, $^2$P$_{3/2}$) transition is resonantly excited by a pump pulse. The transient excited state population is probed "in
real time" on the (5p - (8s, 6d)) transitions with a time-delayed ultrashort
pulse (at 610 nm). The laser system is based on a conventional Ti:
Sapphire laser with chirped pulse amplification (Amplitude Technologies) which supplies
$ 3\, {\rm mJ}$ - $60 \, {\rm fs}$ - $803 \, {\rm nm}$ pulses.  A
fraction of the output is used as the pump pulse.  Another fraction seeds a home made
Non-collinear Optical Parametric Amplifier (NOPA) \cite{nopa-dazzler-apb05} compressed using
double pass silica prisms, which delivers pulses of a few
microJoule, 40 fs -FWHM pulse intensity, centered around 610 nm. The
pump pulse can be shaped with a programmable pulse-shaper. It is recombined with the probe pulse and
sent into a sealed Rubidium cell. The pump-probe signal is detected
by monitoring the fluorescence at 420 nm due to the radiative
cascade (8s,~6d)~-~6p~-~5s collected by a photomultiplier tube as a
function of the pump-probe delay $\tau$. The pulse shaping device is
a 4f set-up composed of one pair each of reflective gratings and
cylindrical mirrors. Its active elements -two 640 pixels liquid
crystal masks- are installed in the common focal plane of both
mirrors. This provides high resolution pulse shaping in phase and
amplitude \cite{pulseshaperRSI04}.

\section{Results and discussion}

\begin{figure}[!ht]
\begin{center}
\epsfig{figure=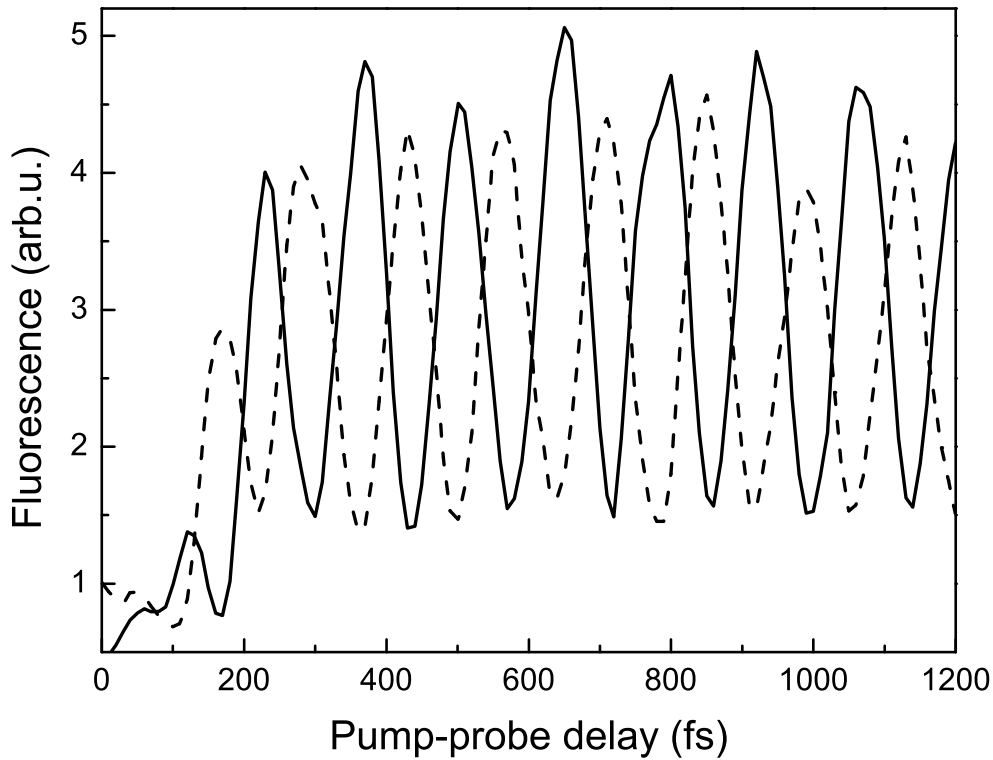,width=0.9\linewidth}
\caption[Experiment]{Experimental results. Fluorescence observed as a function of the pump-probe delay $\tau$. The dashed line represents the population excited by a FT pulse and the solid line represents the shaped case. The origin of the delay time is arbitrary.} \label{Experiment}
\end{center}
\end{figure}

The experimental pump-probe signals are displayed on Figure \ref{Experiment} for the FT and $\pi$ phase jump cases. The time origin is arbitrary. The two oscillations are shifted by $\pi$. In the FT limited case, the first maximum is partially reduced because the pulse duration (60 fs) is only slightly smaller than the oscillation period (140 fs). Therefore, part of the excited population starts to leave the bright state before the end of the pulse. A full maximum is thus only reached after one complete oscillation. With the shaped pulse, the long term dynamics still presents an oscillation which is now shifted by half a period. The first maximum occurs after half a period when the population is maximum in the bright state. A small maximum is also present during the pulse.

The precession dynamics can therefore be clearly controlled. Similarly to what was demonstrated on $\textrm{Li}_2$ \cite{Amitay01shift}, the dynamics could be shifted by an arbitrary offset $T_{shift}$ by applying a relative phase $\theta_{shift} = \Delta \omega \, T_{shift}$. This approach could also be extended to a multilevel system such as a diatomic molecule. For instance, a wave packet could be created on the opposite side of the Franck-Condon point \cite{Garraway98} by applying sign inversions at the frequency corresponding to every other vibrational state. This could be achieved while keeping at a minimal value the population at the Franck-Condon point \cite{Meier_I2shift}.

\section{Conclusion}

As a conclusion, we have shown that the dynamics of spin-orbit precession can be manipulated by applying a sign-inversion at one of the frequencies. This work can be extended in several directions. For instance, with a photoionizing probe, the photoelectron angular distribution \cite{Bouchene01} and its temporal evolution could be controlled in such a way. In more complex systems such as molecular vibration, it should be possible to achieve a significant population of a non Franck-Condon state while keeping at a low level the population of the Franck-Condon state.

Elsa Baynard and St\'ephane Faure are acknowledged for their technical help. We enjoyed stimulating discussions with Chris Meier. This work has been partly supported by the ANR-Programme non th\'ematique (Contract ANR - 06-BLAN-0004).

$^*$ Also at Institut Universitaire de France

\bibliography{bib_control}
\bibliographystyle{apsrev}

\end{document}